\documentclass[twocolumn]{revtex4}
\usepackage{graphicx}
\usepackage{amsmath}
\usepackage{amssymb}
\usepackage{bm}

\begin{document}
\renewcommand{\vec}[1]{\bm{#1}}
%
\title{Twisted Singlet in Semiconductor Artificial Molecules}

\author{Hiroshi Imamura}
\affiliation{Graduate School of Information Sciences, Tohoku University,
  Sendai 980-8579, Japan}
%

\begin{abstract}
The spin-configuration of semiconductor
artificial molecules consisting of quantum dots and spin field effect
transistors is studied theoretically.
We find that the antiferromagnetic spin configuration can be changed
to the ferromagnetic one by applying a gate voltage
to the spin field effect transistors.  We show that 
the square-norm of the total spin of an artificial molecule oscillates
with the twist angle $\theta$ and has maxima at $\theta=$ an odd integer
times $\pi$.  We also show that the square-norm of the total spin of
the ring-shaped artificial molecule changes drastically at certain
values of $\theta$ where the lowest two energy levels cross each other.
\end{abstract} 

\maketitle
%
Nano-spintronics is an emerging research field of physics and
engineering focused on spin-degrees of freedom of electrons confined
in nano-structures \cite{maekawa,awschalom}.
One attractive nano-spintronics device is the
spin field effect transistor (spin-FET) proposed by Datta and
Das\cite{datta1990}, where the
precession of the spin of a conduction electron is dominated by
the Rashba spin-orbit (RSO) interaction\cite{rashba1960,bychkov1984}.
The strength of the RSO interaction and therefore the precession of spin
can be controlled by applying a gate voltage.  The RSO interaction
in semiconductor nano-structures has been extensively studied both
theoretically\cite{datta1990,nitta1999,matsuyama2000,molenkamp2001,pareek2002}
and experimentally\cite{nitta1997,dirk2000,koga2002}.

Semiconductor quantum dot is a basic element of current nanotechnology
and is often considered as an artificial atom\cite{ashoori1996}.  
We are able to construct an artificial molecule (AM) of quantum dots by
connecting the dots either laterally\cite{waugh1995,blick1998} or
vertically\cite{austing1997,pi2001}.   It is important to study the 
spin configuration of electrons in such semiconductor AMs.
The AMs can be described by the Hubbard
Hamiltonian\cite{hubbard1963,sugiura1990}, which has bee extensively 
studied in the context of strongly correlated electron
system\cite{lieb1968,lieb1989,lieb1995,callaway1990}.  
Lieb proved that the ground state of the Hubbard model with a 
bipartite lattice and a half filled band has the antiferromagnetic
spin configuration\cite{lieb1989}.  Lieb's theorem can not be
applied to the lattice with odd number of sites.
However, as we shall show later by using the exact diagonalization
method, the linear-shaped and the ring-shaped one-dimensional lattices
with $N= 3$ and 5 sites also have the antiferromagnetic ground states.
Thus we can say that the ground state of the semiconductor AM with a small
number of quantum dots prefers the antiferromagnetic spin configuration
as long as the system is described by the usual half-filled Hubbard model.

\begin{figure}[t]
  \includegraphics[width=0.95\columnwidth]{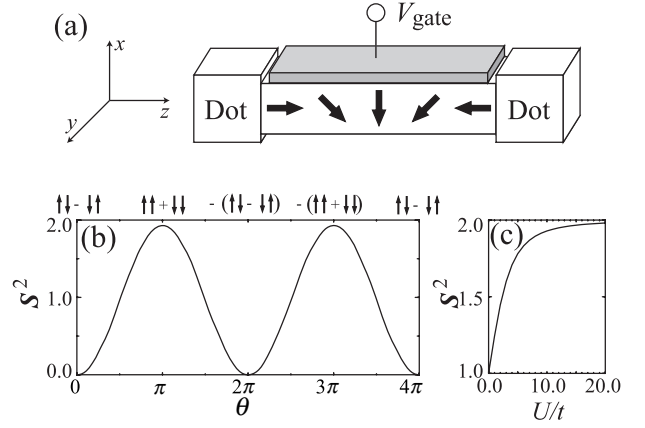}
  \caption{(a) Two quantum dots are connected via a spin-FET.
    The arrows represent the spin of conduction electrons.  The
    gate-voltage is applied to control the strength of the RSO
    interaction.  (b) The square-norm of the total spin $\vec{S}^{2}$
    of the ground state of the two-dot AM is plotted against the twist
    angle $\theta$.
    The  on-site Coulomb energy is assumed to be
    $U/t=10$. The dominant  component of the spin-wavefunction in the
    real spin space is shown on the top.    
    (c) The square-norm $\vec{S}^{2}$ is plotted against the on-site
    Coulomb energy $U/t$.
}
  \label{fig:basic}
\end{figure}

However, if the dots are coupled via spin-FETs as
shown in Fig.\ref{fig:basic} (a) we can control the spin configuration
of the AM by applying a gate voltage to the spin-FETs. 
The key concept is ``twisted
singlet''.  The spin-FET can twist the singlet
spin-wavefunction, which is the essence
of the antiferromagnetism, in the spin space.  If we set the twist
angle to an odd integer times $\pi$, we have the triplet
spin-wavefunction instead of the singlet one.

In this letter, we theoretically study the spin configuration of AMs
consisting of quantum dots and spin-FETs and show that the spin
configuration of such an AM can be controlled by the RSO interaction.
Both linear-shaped and ring-shaped one-dimensional AMs are considered.
It is shown that the square-norm of the total spin of the AM
oscillates with the twist angle $\theta$ and has maxima at
$\theta=$ an odd integer times $\pi$.  It is also shown that the
square-norm of the total spin of the ring-shaped AM changes
drastically at certain values of $\theta$ where the lowest two energy
levels cross each other.  These results provide a powerful guiding
principle of making a gate-controllable nano-magnet. 

The RSO interaction in the spin-FET (See Fig. \ref{fig:basic} (a)) is described by the Hamiltonian
\begin{equation}
  H_{SO}=\alpha
  \left[ \sigma_{y} \nabla_{z} + \sigma_{z} \nabla_{y} \right],
  \label{eq:hso1}
\end{equation}
where $\sigma_{y}$ and $\sigma_{z}$ are respectively the $y$ and
$z$-component of the Pauli spin matrix and $\alpha$ represents the
strength of the RSO interaction.
Experimentally, one typically observes values for $\alpha$
on the order of $10^{-11}$ eV m\cite{nitta1997,dirk2000,koga2002}.
We assume that in the spin-FET the wavefunction in the $y$-direction 
has the form $\sin(m\pi y/W)$, where $W$ is the width of the wire and
$m$ is an integer, the expectation value of $\nabla_{y}$ is zero.
Hence we can neglect the last term in Eq. (\ref{eq:hso1}) and we have
\begin{equation}
 H_{SO}=\alpha \sigma_{y} \nabla_{z}.
\end{equation}
The RSO interaction rotates the spin around the $y$-axis.  
The rotation of the spin by angle $\theta$ in the spin-FET can be represented
by the matrix
\begin{equation}
  R(\theta)
  =\begin{pmatrix}
    \cos \frac{\theta}{2} & - \sin \frac{\theta}{2}\\
    \\
    \sin \frac{\theta}{2} & \cos \frac{\theta}{2}
    \end{pmatrix}.
\end{equation}

We consider the one-dimensional AM of quantum dots
coupled via spin-FETs described by the Hamiltonian
\begin{equation}
  H
  =
  -t\sum_{\sigma\sigma^{\prime}}\sum_{\langle i,j \rangle}
  \left\{
    c_{i\sigma}^{\dag}
    R_{\sigma\sigma^{\prime}}(\theta_{ij})
    c_{j\sigma^{\prime}} 
  \right\}
  +
  U\sum_{i}n_{i\uparrow}n_{i\downarrow},
  \label{eq:hamiltonian}
\end{equation}
where $t$ is the hopping matrix elements, $U$ is the on-site
Coulomb energy, $n_{i\sigma}\equiv c_{i\sigma}^{\dag}c_{i\sigma}$ and operators $c_{i\sigma}^{\dag}$ and $c_{i\sigma}$ are,
respectively, the creation and annihilation operators for an electron
of spin $\sigma$ at $i$th quantum dot.  The sum $\langle i,j \rangle$
is restricted to nearest-neighbor sites and $\theta_{ij}$ is the
rotation angle of spin when the electron is transfered from the $i$th
dot to the $j$th one.  
We assume that system is half-filled, i. e., the number of electrons
is the same as that of dots.
Since we are interested in the spin configuration of AMs, the
physical quantity of central importance is the square-norm of the
total spin defined as 
\begin{equation}
\vec{S}^{2}
  =   \left(S^{z}\right)^{2}
  + \frac{1}{2}\left(S^{+} S^{-} + S^{-} S^{+}\right),
  \label{eq:spin}
\end{equation}
where 
\begin{equation}
  \begin{split}
&  S^{z}=\frac{1}{2}\sum_{i}
  \left(n_{i\uparrow} - n_{i\downarrow}\right)\\
& S^{+}=\sum_{i}c_{i\uparrow}^{\dag}c_{i\downarrow}, \ \ S^{-}=\sum_{i}c_{i\downarrow}^{\dag}c_{i\uparrow}.
\end{split}
\end{equation}

%
Without RSO interaction, Eq. (\ref{eq:hamiltonian}) reduces to 
the usual Hubbard Hamiltonian
\begin{equation}
  H_{t} 
  =
  -t\sum_{\sigma}\sum_{\langle i,j\rangle}
  c_{i\sigma}^{\dag}c_{j\sigma}
  + U\sum_{i} n_{i\uparrow}n_{j\downarrow}.
  \label{eq:norashba}
\end{equation}
The basic properties of the Hubbard Hamiltonian with a finite number
of sites and a half filled band have been investigated intensively by
several authors\cite{lieb1968,lieb1989,lieb1995,callaway1990}.  
The Hamiltonian is commutable with the operator $\vec{S}^{2}$ and the
total spin $S$, $\vec{S}^{2}=S(S+1)$, is the conserved quantity.
According to Lieb's theorem \cite{lieb1989}, the total spin of the
ground state of the bipartite AM of even number of quantum dots
is $S=0$ without RSO interaction.  For the AMs with $N=3$ and 5 
quantum dots, we numerically diagonalize the Hamiltonian
given by Eq. (\ref{eq:norashba}) and confirm that the ground states of
the linear-shaped and the ring-shaped AMs with $N=3$ and 5 have the
antiferromagnetic spin configuration with $S=1/2$.  Without RSO
interaction, all AMs we consider here have antiferromagnetic ground
state with $S=0$ or 1/2.

Let us first consider the simplest AM, two quantum
dots connected via a spin-FET, shown in Fig. \ref{fig:basic} (a).
The Hamiltonian can be easily diagonalized and the energy spectrum is
independent of the angle $\theta_{12}$.
The ground state energy is given by 
\begin{equation}
  E=\frac{1}{2}\left(U-\sqrt{16t^{2}+U^{2}}\right).
\end{equation}
The corresponding spin-wavefunction of the ground state is 
\begin{equation}
\begin{split}
  \psi
  =C 
  \Biggl[
      &\cos\left(\frac{\theta}{2}\right)
      \left(
      |\uparrow,\downarrow\rangle
      -|\downarrow,\uparrow\rangle
    \right) \\
    &+
      \sin\left(\frac{\theta}{2}\right)
    \left(
      |\uparrow,\uparrow\rangle
      +|\downarrow,\downarrow\rangle
    \right)  \\
    &+
    \frac{4 t }{U+\sqrt{16 t^{2} + U^{2}}}
    \left(
      |\uparrow\downarrow,0\rangle
      +|0,\uparrow\downarrow\rangle
    \right)
    \Biggr]
\end{split}
\label{eq:wf}
\end{equation}
where $C$ is the normalization constant, $\theta\equiv\theta_{12}$, 
$|\sigma,\sigma^{\prime}\rangle$ represents the basis where the left dot
has an electron with spin $\sigma$  and the right dot has an electron
with spin $\sigma^{\prime}$, and $|\uparrow\downarrow,0\rangle
(|0,\uparrow\downarrow\rangle)$ represents the basis where the
left(right) dot has two electrons with spin $\uparrow$ and
$\downarrow$.

From Eqs. (\ref{eq:spin}) and (\ref{eq:wf}), the square-norm of the
total spin of the ground state is obtained as
\begin{equation}
  \vec{S}^{2}
  =2\sin^{2}\left(\frac{\theta}{2}\right)
  \frac{
    \left(U+\sqrt{16 t^{2} + U^{2}}\right)^{2}}{16 t^{2} +
    \left(U+\sqrt{16 t^{2} + U^{2}}\right)^{2}}.
\end{equation}
As shown in Fig. \ref{fig:basic} (b), the square-norm  $\vec{S}^{2}$
oscillates with a period of $2\pi$ and has maxima at $\theta=\pi$ and
$3\pi$.  The maximum value of the square-norm $\vec{S}^{2}$ is slightly
smaller than two since the ground state consists not only of the
nearest-neighbor singlet but of the on-site singlet.  As the on-site
Coulomb energy $U/t$ increases, the occupation probability of the
on-site singlet decreases and the maximum value of
$\vec{S}^{2}$ increases as shown in Fig. \ref{fig:basic} (c).

These results for two-dot AM can be easily understood by introducing 
the following local gauge transformation in the spin space:
\begin{equation}
\begin{split}
&c_{2\sigma}^{\dag}\mapsto 
c_{2\sigma}^{\dag}(\theta) \equiv
\sum_{\sigma^{\prime}}R_{\sigma\sigma^{\prime}}(\theta)
c_{2\sigma^{\prime}}^{\dag}\\ 
&c_{2\sigma}\mapsto
c_{2\sigma}(\theta) \equiv
\sum_{\sigma^{\prime}}R_{\sigma\sigma^{\prime}}(\theta)
c_{2\sigma^{\prime}}
\end{split}.
\label{eq:trns}
\end{equation}
The creation and annihilation operators $c_{2\sigma}^{\dag}(\theta)$ and
$c_{2\sigma}(\theta)$ satisfy the usual fermion anti-commutation relations.
The Hamiltonian given by Eq. (\ref{eq:hamiltonian}) is expressed as
\begin{equation}
  \begin{split}
 H
 =
 &-t\sum_{\sigma}\left\{c_{1\sigma}^{\dag}
    c_{2\sigma}(\theta)
    + c_{2\sigma}^{\dag}(\theta) c_{1\sigma}
  \right\}\\
  &
  +
  U\left\{n_{1\uparrow}n_{1\downarrow} +
  n_{2\uparrow}(\theta)n_{2\downarrow}(\theta)\right\},
\end{split}
\label{eq:gt1}
\end{equation}
where $n_{2\sigma}(\theta)\equiv c_{2}^{\dag}(\theta)c_{2}(\theta)$.
Therefore, the energy spectrum is invariant under the gauge
transformation defined by Eq. (\ref{eq:trns}) and is
independent of the angle $\theta$.  The above gauge transformation
corresponds to
the twist of the spin quantization axis of the 2nd dot by angle $\theta$.
In such a $\theta$-twisted spin space, the Hamiltonian is identical
to Eq. (\ref{eq:norashba}).
The spin operators in the $\theta$-twisted spin space are given by
\begin{equation}
\begin{split}
& \vec{S}^{2}(\theta)
  \!=\! \left(S^{z}(\theta)\right)^{2}
  \!+\! \frac{1}{2}\left(S^{+}(\theta)S^{-}(\theta) \!+\!
  S^{-}(\theta)S^{+}(\theta)\right)\\
&  S^{z}(\theta)=\frac{1}{2}
  \left(n_{1\uparrow}-n_{1\downarrow} + n_{2\uparrow}(\theta) -
    n_{2\downarrow}(\theta)\right)\\
&S^{+} = c_{1\uparrow}^{\dag}c_{1\downarrow} +
c_{2\uparrow}^{\dag}(\theta)c_{2\downarrow}(\theta)\\
&S^{-} = c_{1\downarrow}^{\dag}c_{1\uparrow} +
c_{2\downarrow}^{\dag}(\theta)c_{2\uparrow}(\theta).
\end{split}
\end{equation}
The total spin in the $\theta$-twisted spin space $S(\theta)$ ,
$\vec{S}^{2}(\theta)=S(\theta)(S(\theta)+1)$, is the conserved
quantity and the ground state has $S(\theta)=0$.

The $\pi$-rotation operator $R(\pi)$ maps $| \uparrow \rangle$ to $|\downarrow
\rangle$ while $| \downarrow \rangle$ to $-|\uparrow\rangle$.
Therefore, the singlet in the $\pi$-twisted spin space 
$|\uparrow,\downarrow \rangle - | \downarrow,\uparrow \rangle$, which
is the dominant component of the ground state of $S(\theta)=0$, is in fact
the triplet $| \uparrow, \uparrow \rangle + | \downarrow, \downarrow \rangle$
in the real spin space.  The dominant
component of the spin-wavefunction in the real spin space, changes from
singlet $\rightarrow$ triplet $\rightarrow$ singlet $\rightarrow$
triplet as shown on the top of Fig. \ref{fig:basic} (b).

\begin{figure}[t]
  \includegraphics[width=0.95\columnwidth]{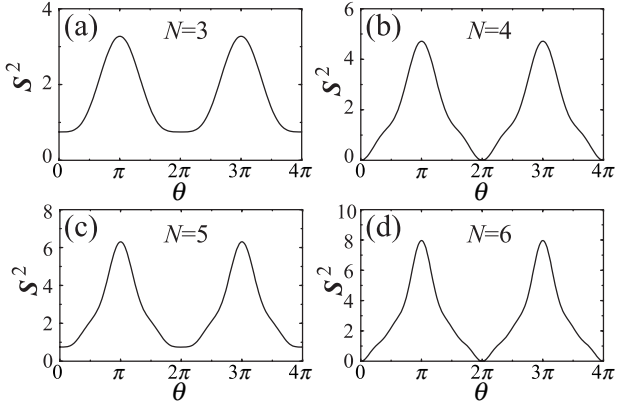}
    \caption{
      The square-norm of the total spin $\vec{S}^{2}$ of the ground states
      of linear-shaped AMs with N=3, 4, 5  and 6 are plotted against the
      twist angle $\theta$  in panels (a), (b), (c) and (d), respectively.  
      The on-site Coulomb energy is $U/t=10$ and the twist angle takes the
      same value$\theta$ for all spin-FETs.}
    \label{fig:line}
\end{figure}

In Figs. \ref{fig:line} (a)-(d), we show the square-norm of the ground state
$\vec{S}^{2}$ for the linear-shaped AM consisting of $N=3, 4, 5$, and
6 quantum dots.  The energy levels and corresponding spin-wavefunctions
are obtained by using the exact diagonalization method.
We assume that the on-site Coulomb energy is $U/t=10$ and the twist
angle takes the same value $\theta$ for all spin-FETs.
Without RSO interaction, $\theta=0$, the ground state has the
antiferromagnetic spin configuration with $S=0$ or 1/2.
The square-norm $\vec{S}^{2}$ is again an oscillating function of 
$\theta$ with a period of 2$\pi$.  However the shoulders appear at
$\theta = \pi/2, 3\pi/2,5\pi/2$ and $7\pi/2$ for $N = 4, 5$ and 6.
The shoulders correspond to the formation of the next-nearest-neighbor
twisted singlet.  For $N = 6$ we have other shoulder structures at
$\theta = \pi/3, 5\pi/3, 7\pi/3$, and $11\pi/3$ corresponding to the
next-next-nearest-neighbor twisted singlet.

\begin{figure}[t]
  \includegraphics[width=0.95\columnwidth]{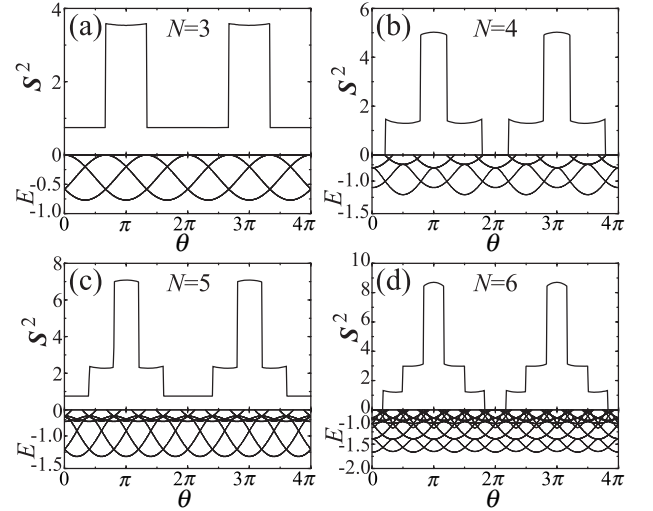}
    \caption{
      The square-norm of the total spin $\vec{S}^{2}$ of the ground states
      and energy levels of ring-shaped AMs with $N=3, 4, 5$
      and 6 are plotted against the twist angle $\theta$ in panels (a), (b),
      (c) and (d), respectively.  The energy is normalized by the hopping
      matrix element $t$. The on-site Coulomb energy is $U/t=10$ and the
      twist angle takes the same value $\theta$ for all spin-FETs.
}
  \label{fig:ring}
\end{figure}

The operators in the twisted spin space are obtained by using the
following local gauge transformation in the spin space:
\begin{equation}
\begin{split}
& c_{i\sigma}^{\dag}\mapsto
c_{i\sigma}^{\dag}(\theta_{i}) \equiv
\sum_{\sigma^{\prime}}R_{\sigma\sigma^{\prime}}(\theta_{i})
c_{i\sigma^{\prime}}^{\dag}\\
& c_{i\sigma}\mapsto
c_{i\sigma}(\theta_{i}) \equiv
\sum_{\sigma^{\prime}}R_{\sigma\sigma^{\prime}}(\theta_{i})
c_{i\sigma^{\prime}}
\end{split}
\label{eq:gtgen}
\end{equation}
The Hamiltonian and spin operators in the twisted spin space are
obtained by replacing operators $c_{i\sigma}^{\dag}$ and $c_{i\sigma}$
in Eqs.(\ref{eq:hamiltonian}) and (\ref{eq:spin}) by
$c_{i\sigma}^{\dag}(\theta_{i})$ and $c_{i\sigma}(\theta_{i})$ as
\begin{equation}
  H_{t} 
  \!=\!\!
  -t\sum_{\sigma}\sum_{\langle i,j\rangle}
  c_{i\sigma}^{\dag}(\theta_{i})c_{j\sigma}(\theta_{j})
  \!+\! U \!\sum_{i} n_{i\uparrow}(\theta_{i})n_{i\downarrow}(\theta_{i}),
\end{equation}
where $n_{i\sigma}(\theta_{i})\equiv
c_{i\sigma}^{\dag}(\theta_{i})c_{i\sigma}(\theta_{i})$.  The energy spectrum is independent of the twist angle.  The spin
operators in the twisted spin-space are given by
\begin{equation}
\vec{S}^{2}(\theta)
  \!=\!   \left(S^{z}(\theta)\right)^{2}
  \!+\! \frac{1}{2}\!\left(S^{+}(\theta) S^{-}(\theta) + S^{-}(\theta)
  S^{+}(\theta)\right),
\end{equation}
where 
\begin{equation}
  \begin{split}
&  S^{z}(\theta)=\frac{1}{2}\sum_{i}
  \left(n_{i\uparrow}(\theta_{i}) - n_{i\downarrow}(\theta_{i})\right)\\
& S^{+} = \sum_{i} c_{i\uparrow}^{\dag}(\theta_{i})
  c_{i\downarrow}(\theta_{i}), \ \
  S^{-} = \sum_{i} c_{i\downarrow}^{\dag}(\theta_{i})
  c_{i\uparrow}(\theta_{i}).
\end{split}
\end{equation}
The ground state has the antiferromagnetic spin configuration  in the
twisted spin space with $S(\theta)=$ 0 or 1/2.

On the contrary to the linear-shaped AM, the square-norm
$\vec{S}^{2}$ of the ring-shaped AM is not a smooth function of the
twist angle.
In Figs. \ref{fig:ring} (a)-(d), we show the square-norm
$\vec{S}^{2}$ and the energy levels for the ring-shaped AMs
with $N=3, 4, 5$ and 6 dots obtained by using the exact
diagonalization method.  We assume that the on-site Coulomb energy is
$U/t=10$ and the twist angle takes the same value $\theta$ for all
spin-FETs.
Without RSO interaction, $\theta=0$, the ground state has the
antiferromagnetic spin configuration with $S=0$ or 1/2.
Due to the  boundary condition of the ring-shaped AMs,  it is
difficult to twist the spin configuration by an arbitrary angle.
Therefore, the square-norm $\vec{S}^{2}$  jumps at a certain
value of angle $\theta$ where the lowest two energy levels cross each
other as shown in Figs. \ref{fig:ring} (a)-(b).  

In conclusion, we theoretically study the spin-configuration of
artificial molecules which consists of quantum dots and spin field
effect transistors. We show that we can change the antiferromagnetic
spin configuration to the ferromagnetic one by applying a gate voltage
to the spin field effect transistors. We note that the approach of
twisting the singlet by using the spin field effect transistor is a
very powerful tool for making the ferromagnetic spin configuration in
non-magnetic semiconductor nano-structures.  The
periodic oscillation of the square-norm of the total
spin in artificial molecules against the twist angle can be applied
to the future nano-spintronics devices, for example, a
gate-controllable nano-magnet.

This work was supported by MEXT, Grant-in-Aid for Scientific Research
on the Priority Area "Semiconductor Nanospintronics" No. 14076204
,Grant-in-Aid for Encouragement of Young
Scientists, No. 13740197 and Grant-in-Aid for Scientific
Research (C), No. 14540321.

\end{document}